\documentclass[preprint2]{aastex}

\newcommand{\myemail}{klimek@physics.utexas.edu}
\newcommand {\hii}{H\,\textsc{ii}}
\newcommand {\hi}{H\,\textsc{i}}
\newcommand {\ha}{H$\alpha$}
\newcommand {\sii}{[S\,\textsc{ii}]}

\newcommand {\oiii}{[O\,\textsc{iii}]}
\newcommand{\heii}{He\,\textsc{ii}}

\newcommand {\kms}{km~s$^{-1}$}
\newcommand {\rosat}{{\it ROSAT\/}}
\newcommand {\xmm}{{\it XMM-Newton\/}}

\newcommand {\Msun}{$M_{\odot}$}
\newcommand{\Vhot}{V_{\mathrm{hot}}}
\newcommand{\Vshell}{V_{\mathrm{shell}}}
\newcommand {\NH}{\mbox{$N_{\rm H}$}}        
\def\farcm  {\hbox{$.\mkern-4mu^\prime$}}

\newbox\grsign \setbox\grsign=\hbox{$>$} \newdimen\grdimen \grdimen=\ht\grsign
\newbox\simlessbox \newbox\simgreatbox
\setbox\simgreatbox=\hbox{\raise.5ex\hbox{$>$}\llap
     {\lower.5ex\hbox{$\sim$}}}\ht1=\grdimen\dp1=0pt
\setbox\simlessbox=\hbox{\raise.5ex\hbox{$<$}\llap
     {\lower.5ex\hbox{$\sim$}}}\ht2=\grdimen\dp2=0pt

\newcommand {\snra}{SNR0449--6921} 
\newcommand {\snrb}{SNR0506--6542} 
\newcommand {\snrc}{SNR0537--6628}

\slugcomment{Published in ApJ, 725, 2281-2289}

\shorttitle{X-ray Investigation of Three SNRs in the LMC}
\shortauthors{Klimek et al.}

\begin{document}


\title{An X-ray Investigation of Three Supernova Remnants in the Large Magellanic Cloud}


\author{M. D. Klimek\altaffilmark{1}, S. D. Points, R. C. Smith}
\affil{Cerro Tololo Inter-American Observatory, National Optical Astronomy Observatory, Casilla 603, La Serena, Chile}
\email{spoints@ctio.noao.edu, csmith@ctio.noao.edu}

\author{R. L. Shelton}
\affil{ Department of Physics and Astronomy, University of Georgia, Athens, GA 30602}
\email{rls@physast.uga.edu}

\and

\author{R. Williams}
\affil{Columbus State University, Coca-Cola Space Science Center, 701 Front Ave., Columbus, GA, 31901}
\email{rosanina@ccssc.org}


\altaffiltext{1}{Present email address: \myemail}


\begin{abstract}

We have investigated three SNRs in the LMC using multi-wavelength data.  These
SNRs are generally fainter than the known sample (see \S4) and may represent a previously
missed population.  One of our SNRs is the second LMC remnant analyzed which is
larger than any Galactic remnant for which a definite size has been established.
The analysis of such a large remnant contributes to the understanding of the 
population of highly evolved SNRs.

We have obtained X-ray images and spectra of three of these recently 
identified SNRs using the \xmm\ observatory.  These data, in conjunction 
with pre-existing optical emission-line images and spectra, 
were used to determine the physical conditions of the
optical- and X-ray--emitting gas in the SNRs.  We 
have compared the morphologies of the SNRs in the different wavebands.  The 
physical properties of the warm ionized shell were determined from the 
\ha\ surface brightness and the SNR expansion velocity.  The X-ray 
spectra were fit with a thermal plasma model and the physical 
conditions of the hot gas were derived from the model fits.  
Finally, we have compared our observations with simulations of SNR evolution.
\end{abstract}





\section{Introduction}

Supernovae (SNe) and their remnants are the driving force behind the
dynamic interstellar medium (ISM) of our Galaxy and others.  The study
of the energetics and evolution of supernova remnants (SNRs) thus forms
the foundation for developing a complete understanding of the complex
structure and evolution of the ISM in gas-rich galaxies not unduly affected by external events such as galaxy-galaxy
collisions..  However, in order to form an accurate picture of how SNRs influence the ISM, it is necessary to have a complete and unbiased sample of SNRs in a galaxy.  This is very difficult within the Milky Way as dust absorption and confusion along the line-of-sight hinders our ability to detect and determine the physical conditions of Galactic SNRs.  Although it is possible to obtain a more complete sample of SNRs in more distant galaxies, their distance hinders the investigation of their physical conditions in detail.

The Large Magellanic Cloud (LMC) offers the ideal laboratory
for the study of a large sample of SNRs.  The LMC is well out of the Galactic plane and thus offers low foreground absorption ($A_{V}\sim 0.3$).  Meanwhile, its distance is relatively well known ($\sim$50 kpc; \citet{dibenedetto}, \citet{feast}) and distances within the LMC are small compared with its distance to us so all objects within the LMC can be taken to be at approximately the same distance.  Its proximity also allows us to resolve a great amount of spatial detail (1"$\approx 0.25$~pc).

The SNRs presented in this paper were discovered as regions of shock-ionized 
gas in the Magellanic Cloud Emission Line Survey (MCELS) using the reliable 
\sii/\ha\ diagnostic \citep{fesen}, with SNRs typically having \sii/\ha\ $> 
0.4$ \citep{siihavalue}.  Analysis of archival \rosat\ data and high resolution 
optical spectra confirmed these shells as SNRs (\citet{inprep}, \citet{aas}).
However, the \rosat\ observations are too shallow to provide much detailed 
information on the hot gas within these remnants. Two of these remnants 
have been detected, but not classified as SNRs in the \rosat\ survey of the LMC 
\citep{hp99}, and the Parkes radio survey \citep{filipovic}. 
\citet{blair} seached for FUV emmission from these remnants but failed to detect them.

A thorough investigation of the hot gas is crucial to understanding the properties of the SNR and its interaction with the host galaxy.  SNRs contribute most of the hot gas component of the ISM, and much of the energy originally released in the supernova event is carried as thermal energy in the hot gas.  Moreover, the interaction between the hot interior and the cool shell of shocked material dictates the evolution of the SNR.  We have therefore obtained new X-ray data from the \xmm\ observatory and analyzed these data along with the optical data from MCELS to form a complete picture of these SNRs.

Our observations and analysis techniques are presented in \S2.  In \S3 we give the results of these techniques applied to the individual objects, followed by discussion of those results in \S4.  A brief conclusion is given in \S5.

\section{Observations}

\subsection{X-ray}

\xmm\ is an orbiting X-ray observatory operated by the European Space Agency.  In this study, we use data from the European Photon Imaging Camera (EPIC) MOS CCDs.  These are two identical devices each mounted behind 58 nested Wolter type 1 mirrors.  The EPIC cameras offer the capability to perform extremely sensitive imaging observations over the telescope's field of view of 30 arcmin and in the energy range from 0.15 to 15 keV with moderate spectral ($E/\Delta E \sim 20-50$) and angular resolution (point-spread function: 6 arcsec FWHM).  For further technical discussion, see \citet{turner}.

This study presents new X-ray data from XMM-Newton
on three LMC SNR candidates:   \snra, \snrb, and \snrc.     
 Proposed pointings were chosen to follow up on regions of high \sii/\ha\ from the MCELS data. 
 Three observations are presented in this paper. 
 See Table \ref{obslog} for details on the observations. 

These data were processed with the Extended Source Analysis Software (ESAS) package.  ESAS is a package of perl scripts which call commands from ESA's Science Analysis Software and standalone FORTRAN 77 programs following analysis techniques given by \citet{snowden}.  ESAS automatically filters times of high background contamination and poor event grades.  

The low X-ray surface brightness of these SNRs, in combination with limits of X-ray coverage and resolution from past surveys, explains their non-detection until the present.  We will show that the average X-ray surface brightness of these remnants falls below the lowest levels detected from LMC SNRs included in earlier catalogs.  Given the relatively small number of source counts we are able to collect from the remnants, we need to have a good understanding of the backgrounds so that we can accurately extract source properties from limited data.

After the data had been filtered, we applied the ESAS {\tt cheese} task, which uses SAS tools to identify point sources, and creates a mask file.  We specifically ran the source detection over a band from 0.8 to 8.0 keV in order to identify troublesome nonthermal background sources without falsely masking softer lumpy features within the SNR emission.

\begin{figure}
\epsscale{2}
\plotone{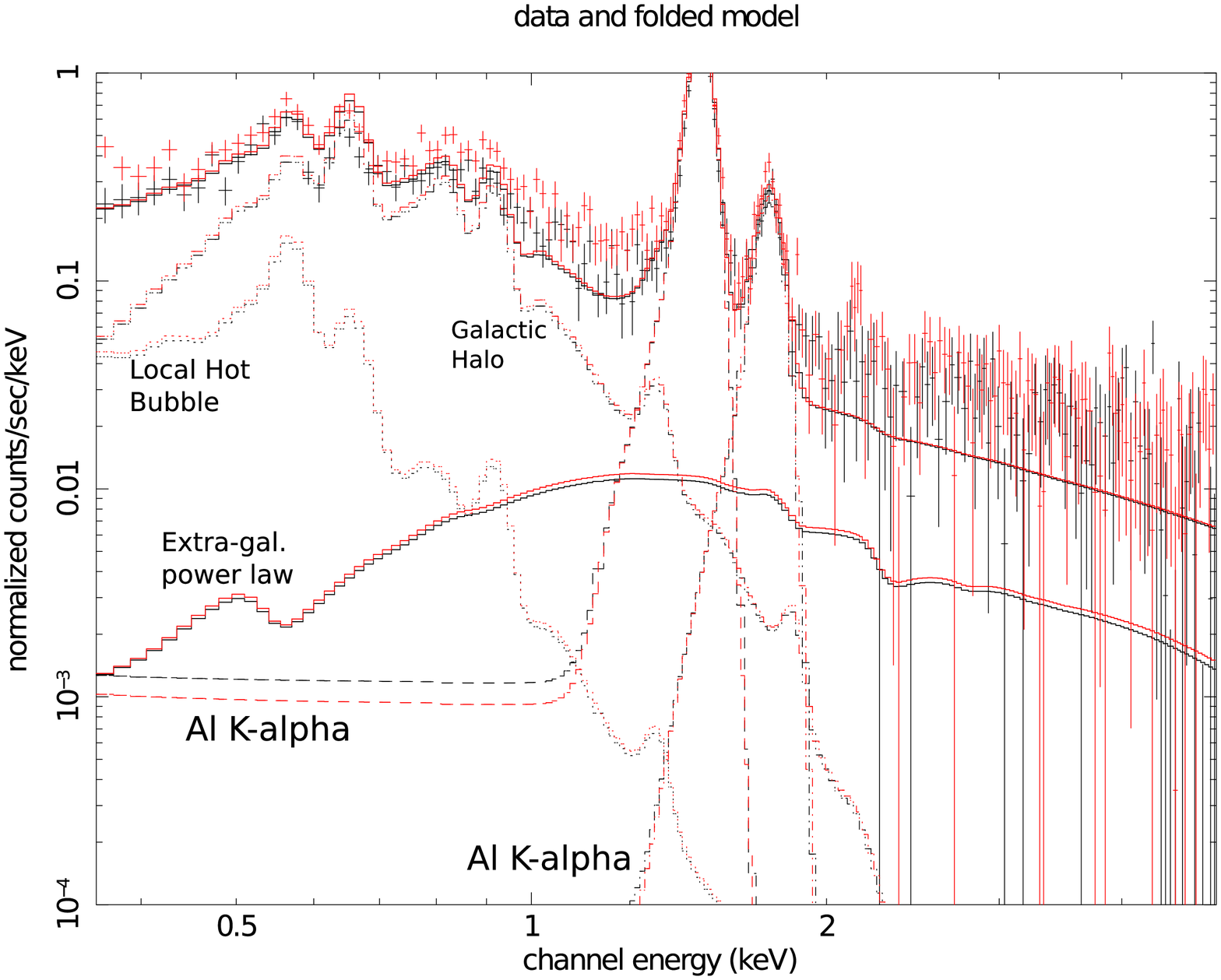}
\caption{The background spectrum for \snrb\ with individual components plotted and labeled.}\label{bgcomponents}
\end{figure}

Using these masks, we then extracted spectra from background regions near the remnants.  We binned these background spectra to 50 counts per channel and fit them with NASA's Xspec package using a model composed of two gaussian components to account for the strong Al and Si K$\alpha$ fluorescent lines near 1.5 and 1.75 keV, respectively, an APEC component for the Local Hot Bubble/Solar Wind Charge Exchange, an absorbed APEC component for the Milky Way's halo, and an absorbed power law component for the nonthermal extragalactic background.  The photon index of the extragalactic power law was fixed at 1.46 \citep{snowden, chen}.  Finally, we also included a power law component which was not folded with the instrumental response to model residual soft proton contamination not removed in the light curve filtering.  The specifics of the background modeling are discussed below for each observation, but aside from the soft proton contamination which varies with time, the background parameters found by this method are consistent between all the observations.  In addition, our temperature for the local hot bubble in \snra\  and \snrc\ are consistent with the results of \citet{henley_shelton}.  An example of our background model is seen in Figure \ref{bgcomponents}.

\begin{figure}
\epsscale{2}
\plotone{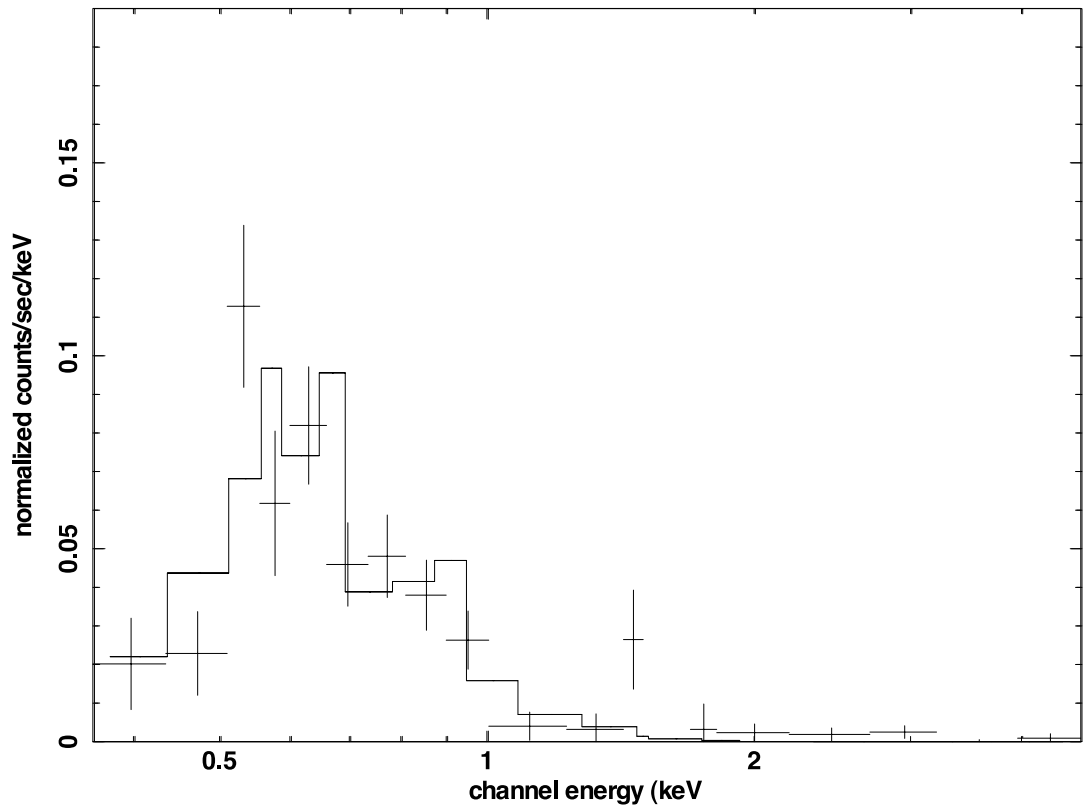}
\caption{Background-subtracted source spectrum for \snrb\ fitted with our source model.}.\label{model}
\end{figure}

Next we extracted spectra of the SNR emission and fit each with the corresponding background model, plus an APEC thermal plasma component, with abundances set to the canonical LMC value of 0.3 solar.  We froze the temperatures, photon indices, column density, and gaussian parameters as determined from the background fit, but allowed the normalizations to vary to account for the change in area between the background and source spectrum extraction regions.  The source spectra generally have few energy bins with significant numbers of counts, but the constraints provided by the background fits allowed us to acheive statistically significant fits for the source parameters.  Fits using the older Raymond-Smith model or the MEKAL model in place of the APEC model gave nearly identical results.  After this initial fitting, we added the additional constraint that the ratio of the normalizations of the absorbed thermal and nonthermal components should remain constant over the field-of-view.  This constraint did not change the best-fit values that we found, but did improve the uncertainties in the fitted values.  An example of the residuals from our background model fitted with the APEC model is seen in Figure \ref{model}.

We compared our fitted values of hydrogen column density with that found in \hi\ surveys.  We used the Leiden/Argentine/Bonn survey of Galactic hydrogen \citep{kalberla} and the ATCA+Parkes survey of LMC hydrogen \citep{kim}.  It has been shown \citep{ab} that above $\NH>5\times10^{20}$~cm$^{-2}$, the total X-ray absorption column increases by about a factor of 2 over the \hi-derived value due to molecular hydrogen.  We must also remember that only part of the LMC column is actually in front of our objects.  Thus we should expect our fitted values to be close to $2\NH^\mathrm{Gal}+\NH^\mathrm{LMC}$, where $\NH^X$ is the value quoted from the respective survey.  In every case, we found that our fitted values of \NH\ are consistent with the survey values.

Finally, we compared the observed X-ray spectrum with simulations.  We simulated each SNR with the aid of a computer code that followed
the hydrodynamics of the plasma on a spherically-symmetric Lagrangian
mesh, simulated the shock front, shock heating, thermal conduction of
heat from the remnant's hot center outwards, radiative cooling by
non-collisional ionizational equilibrium (non-CIE) plasma, and,
when needed, nonthermal pressure due to magnetic fields and cosmic
rays.   The products of our simulations, the
temperature, density, and ionization level as a function of
location within the SNR and SNR age, were used to calculate
the spectra in a non-CIE manner.   The algorithms are described
in detail in \citet{shelton}.  The spectra were then converted to Xspec models with age
as a free parameter.  To accomplish the spectral fitting in Xspec, we used the fit to the source region as described above, and assumed the APEC source component accurately models the SNR plasma emission.  We then deleted the APEC component and replaced it with the simulated spectrum.  We stepped through the spectra at various ages and searched for a minimum in the Chi-squared.  The simulations were run assuming solar abundances.  As a first approximation, we varied the normalization of the resulting spectral model to account for the different abundances of the LMC.

We also used ESAS to produce combined and adaptively smoothed images in three bands (0.35-0.85, 0.85-2.0, and 2.0-8.0 keV) with a kernel of 15 counts.  The ESAS task {\tt proton} uses the soft proton contamination model parameters found in the background fitting to subtract this component from the images.

Given the thermal plasma model normalization $A$ we can calculate the electron density in the hot gas by $n_{{e}}=3.89\times 10^{7} D\sqrt{A/\Vhot }$ where $D$ is the distance to the SNR, $A$ is the normalization of the plasma model component, and $\Vhot$ is the volume of the hot gas in cgs units.  The mass of the hot gas is $M_{\mathrm{hot}}=1.17n_{{e}}m_{{p}}\Vhot $, assuming fully ionized helium and where $m_p$ is the proton mass.  
Given the temperature from our spectral fits, the total thermal energy is then $E=(3/2)NkT=4.60\times 10^{-9}n_{{e}}\Vhot T$, where $T$ is the temperature of the plasma model in keV. We can also find the pressure as $P_{\mathrm{hot}}=nkT=3.05\times 10^{-9}n_{{e}}T$.  These calculated values are listed in Table \ref{snrtable}.  The errors are the one sigma uncertainties estimated by Xspec in the fit process propagated through the above equations.

\subsection{Optical}

Optical images came from the MCELS project \citep{mcels} where these remnants were originally identified.  The observations consist of red and green continuum images and \ha, \sii, and \oiii\ emission line images from the Curtis Schmidt Telescope at the Cerro Tololo Inter-American Observatory (CTIO).  The final images are flux calibrated and rescaled to have 2'' pixels.  The continuum exposures were used to subtract the continuum contribution to the emission line images. 

In addition, we have long-slit, high-dispersion echelle spectra of the SNR
candidates from the CTIO 4m telescope.  The
data have a 24$\mu$m
pixel size that corresponds to 0.082 \AA\ (3.65~\kms) along the
dispersion axis and $\sim 0\farcs 26$ along the sky.  The spatial
coverage along the slit is roughly $3'$, limited by the optics of the
camera.  
Details of the reduction of these data can be found in \citet{inprep}.  

The echelle profile of the \ha\ line is broadened along the dispersion axis both by the inherent instrumental profile ($\sim14$~\kms) and by thermal Doppler broadening ($\sim18$~\kms~ at X-ray temperatures).  To assign an expansion velocity to each remnant, we extract profiles along the dispersion axis where the emission shows the greatest dispersion and measure the peaks of the profile.  

Using the flux-calibrated MCELS images, we measured an \ha\ surface brightness and the average radius $R$ and thickness $\Delta R$ of the shell.  For a uniform spherical shell, the greatest line of sight through the shell is $\mathcal{L}=2\sqrt{R^{2}-(R-\Delta R)^{2}}$.  The measured surface brightness then implies an emission measure $\mathrm{EM}\equiv n^{2}_{e}\mathcal{L}=5\times10^{17}\times \mathrm{SB}$ where SB is the surface brightness in cgs units and arcseconds.
The total mass of the warm ionized shell is $M=1.27n_{{e}}m_{{p}}\Vshell$, where $\Vshell$ is the volume of the shell and assuming singly ionized helium. Expansion velocities $v_\mathrm{exp}$ have been determined from echelle spectra of the \ha\ emission-line, and thus we may determine kinetic energies for the warm shells by $K=Mv^{2}_{\mathrm{exp}}/2$.  If we assume that $T=10^{4}$~K, we can also calculate the pressure in the shell as $P_{\mathrm{shell}}=2n_{{e}}kT=2.76\times 10^{-12}n_{{e}}$.  

In calculating the volume of the shell, we assumed an ellipsoidal geometry, measuring two axes from the projected face of the shell and taking the third line-of-sight axis to be the average of these two.  We then took the uncertainty in the radius along this third axis to be the deviation of the first two axes from their mean.
We also considered the instrumental and Doppler broadening of our echelle spectra as an additional source of error.  Convolving the two profiles, we found an uncertainty of 11 \kms.
Calculated quantities are listed in Table \ref{snrtable}.  
  
The age of the SNR may be estimated using the analytic expressions of the Sedov-Taylor solution for blast wave expansion, in which the shell radius is given as a function of time by $ r(t)=1.17(Et^2/\rho)^{1/5}$, where $E$ is the explosion energy and $\rho$ is the ambient density.  Taking the time derivative and applying the Rankine-Hugoniot conditions for a strong shock to relate the blast wave velocity to the post-shock gas velocity $u$, we find $u(t)=0.351(E/\rho t^3)^{1/5}$.  Solving this system yields $t=0.3(r/u)$.  We then obtain the ages by plugging in the radius observed in the optical images and the velocities from the echelle spectra.  
These procedures apply best to remnants that are still in the Sedov-Taylor phase of
evolution, since post-Sedov-phase remnants continue to expand at a slower rate.  The Sedov-Taylor phase is expected to end when the post-shock gas velocity drops to $\sim190$~km s$^{-1}$.
     Inhomogeneities
in the ambient medium also affect the expansion rate.   In addition, the observed velocities and radii
are found for individual parts of the SNRs and, due to asymmetries in the SNRs may not adequately
represent the whole objects.   All of these effects combine to make the ages estimated from $t = 0.3(r/u)$ approximate.

The simulations described previously also provide an independent numerical estimate of the age.  Alternatively, given the present values of $r$ and $u$, one may solve for the ratio $E/\rho$, or equivalently $E/n$ where $n$ is the ambient number density, and compare this with the value estimated using the X-ray spectral modeling and H$\alpha$ surface brightness techniques as described above.

\section{Results}

\subsection{\snra}

\begin{figure}
\plotone{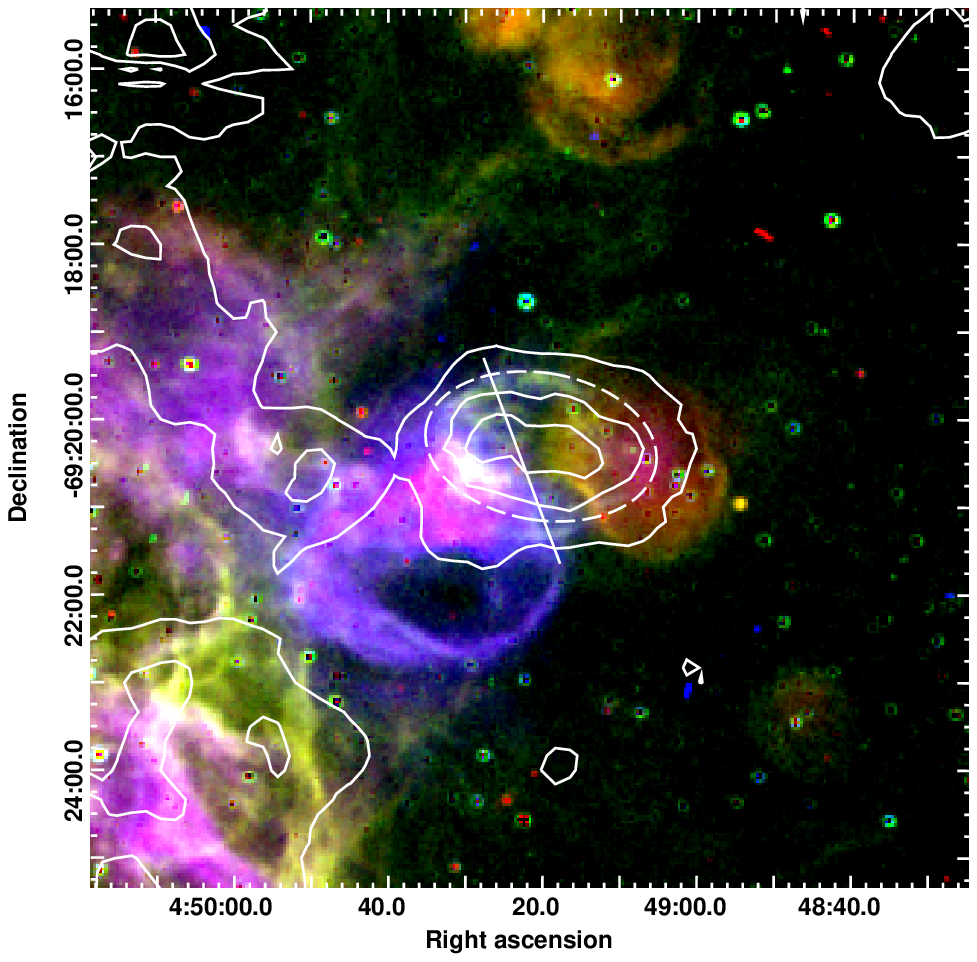}
\caption{Three-color image of \snra\ with X-ray contours overlaid. Red is \ha, green is \sii, and blue is \oiii.  Contours are at levels of 2 counts per second per square degree in the energy band 0.35--0.85 keV.  The dashed ellipse corresponds to the SNR extent considered in the calculations of the X-ray and optical shell properties.  The line corresponds to the radial slice shown in Figure \ref{rad0449}.}\label{fig0449}
\end{figure}

\begin{figure}
\epsscale{1.5}
\plotone{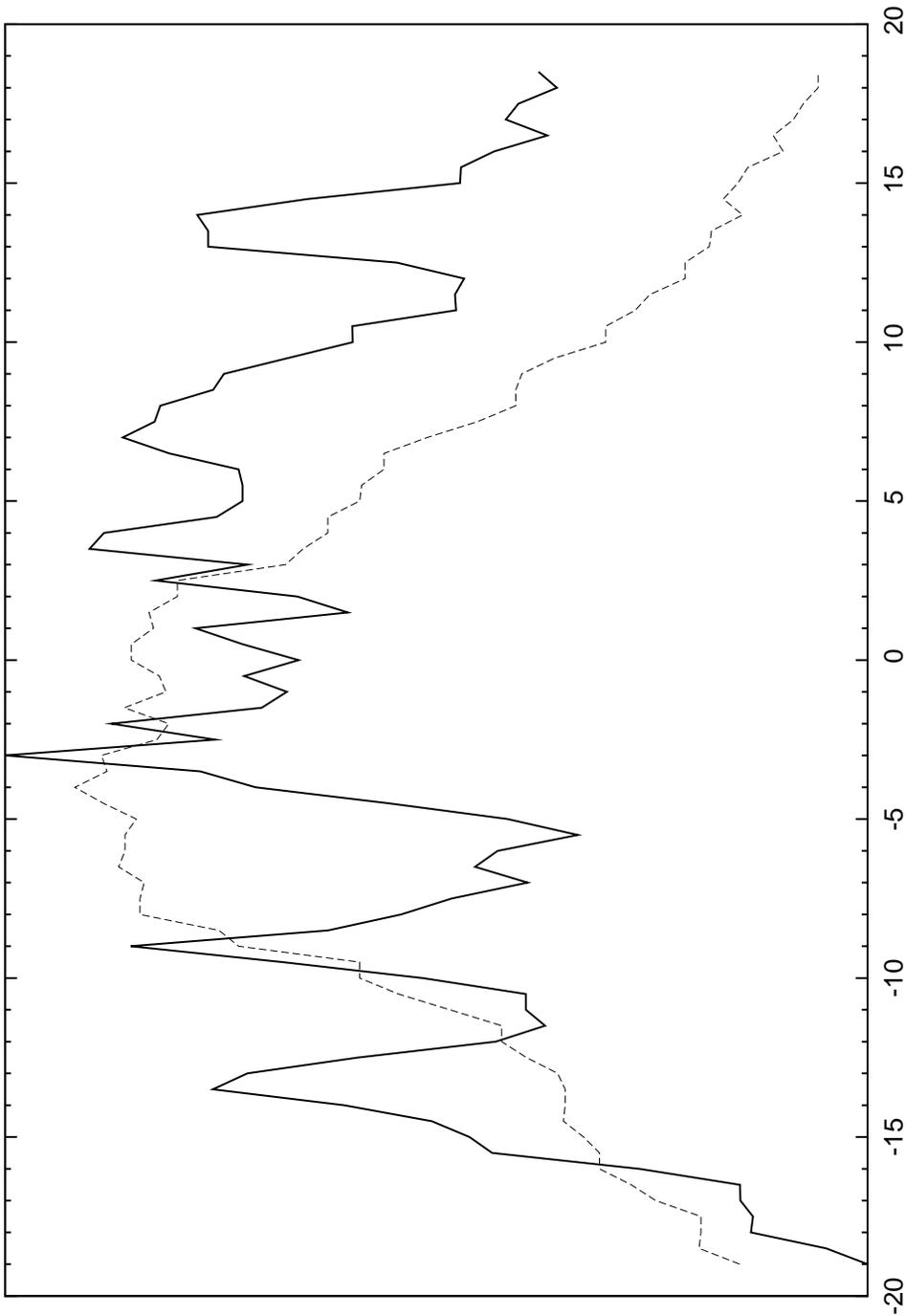}
\caption{Radial profile of \snra.  The solid line represents \ha\ and the dashed line represents 0.35--0.85 keV smoothed X-rays.  X-axis units are parsecs.  The origin corresponds to  04 49 22.74, -69 20 27.28.  The slice was chosen so as to avoid the diffuse emission from the nearby \hii\ region.}\label{rad0449}
\end{figure}

\snra\ shows the simplest structure of the three remnants presented here.  Its well defined \sii\ shell surrounds a smooth elliptical distribution of soft X-ray emission with radii of $1.3'\times0.75'$ or $19.4\times11.25$~pc.
The shell has typical values for \sii/\ha\ of $\sim0.6$, as opposed to maximum values $\sim0.2$ elsewhere.
The bright \ha\ emission to the east of \snra\ comes from the \hii\ region \object[LHA 120-N 79]{N79} and is not directly related to our study.  See Figures \ref{fig0449} and \ref{rad0449}. 

\object[LHA 120-N 79]{N79} corresponds to an area of diffuse soft X-ray emission which covers half the XMM field of view.  We defined a circular source region covering the SNR and then divided the field of view into two background regions corresponding to the area covered by the diffuse X-ray emission and the empty area.  The spectra from the two background regions were fit simultaneously with independent normalizations but other model parameters tied together.  In the region with diffuse X-rays we included an additional absorbed thermal component.

\snra\ shows a clear bow-shaped expansion pattern in the echelle spectrum.  It is difficult to discern a complete corresponding blueshifted pattern.  It may be obscured by the broad band of emission from the surrounding \hii\ region if \snra\ is moving away along our line of sight with respect to the systemic velocity.  Some blueshifted emission is seen along the right side of the spectrum.  Taking this to be the front edge of the shell, we measured a velocity of $\pm70$~\kms.

We found a best-fit temperature for the local hot bubble of $kT=0.11$~keV, and for the Galactic halo of $kT=0.27$~keV.  In the region with the diffuse soft X-rays we found an additional absorbed thermal component with $kT=0.19$~keV.  Best-fit hydrogen column density is 3.7$\times 10^{21}$~cm$^{-2}$.  Full details of the background fit are found in Table \ref{bgtable}.

Unfortunately all but 7 ks of this observation was lost to flaring events.  This combined with the fact that \snra\ is small and faint resulted in too few source counts to constrain the SNR properties, even with the information provided by the background fits.    However, by defining emission above 1.5 keV as hard, we can estimate a hardness ratio of -0.97.  Based on the clear detection of soft X-ray emission confined within a well-defined shell of enhanced \sii/\ha\ optical emission, we identify this candidate as a bona fide SNR.

\subsection{\snrb}

\begin{figure}
\plotone{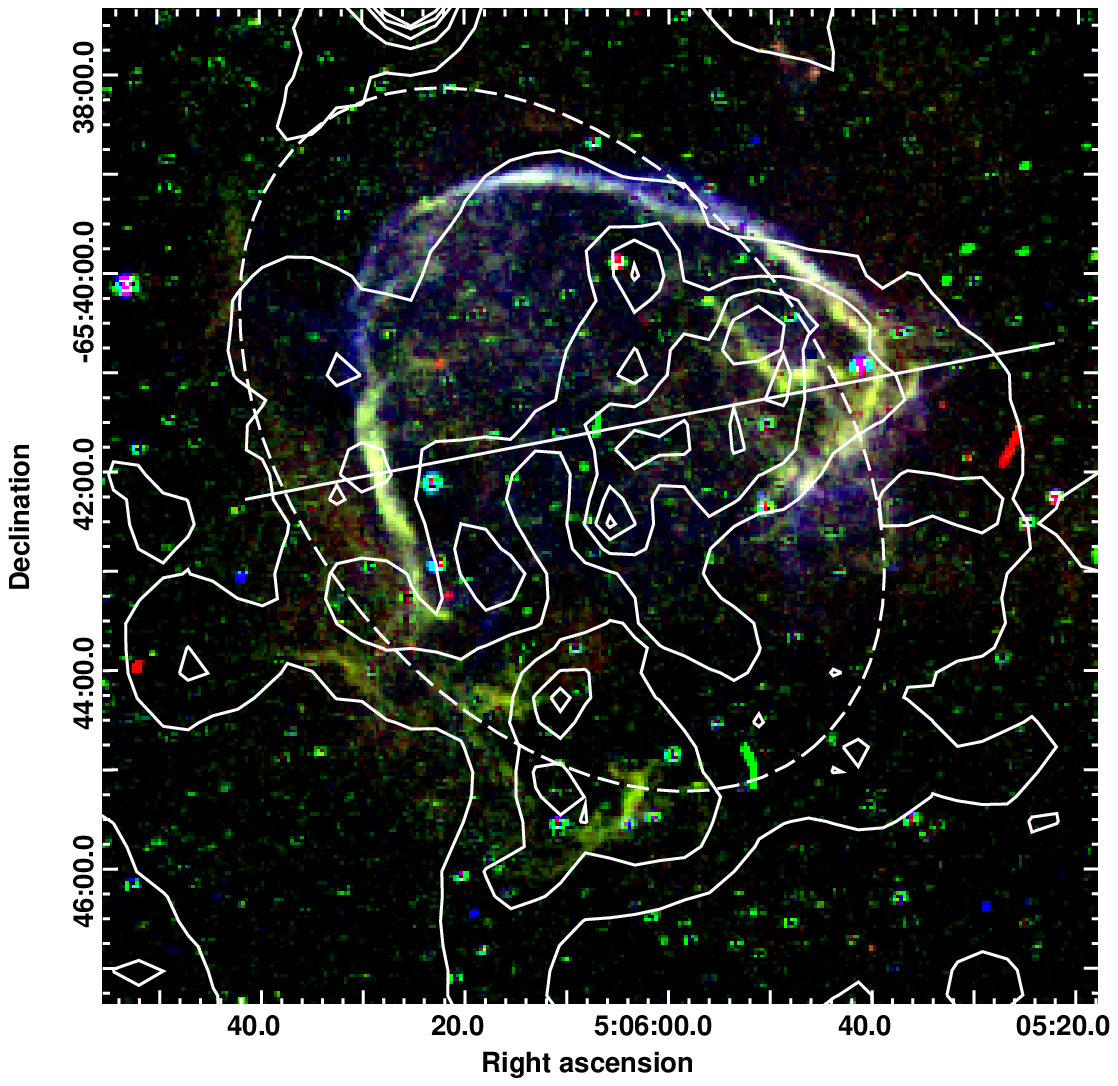}
\caption{Three-color image of \snrb\ with X-ray contours overlaid. Red is \ha, green is \sii, and blue is \oiii.  Contours are at levels of 2 counts per second per square degree in the energy band 0.35--0.85 keV.  The dashed ellipse corresponds to the SNR extent considered in the calculations of the X-ray and optical shell properties.  The line corresponds to the radial slice shown in Figure \ref{rad0506}.}\label{fig0506}
\end{figure}

\begin{figure}
\epsscale{1.5}
\plotone{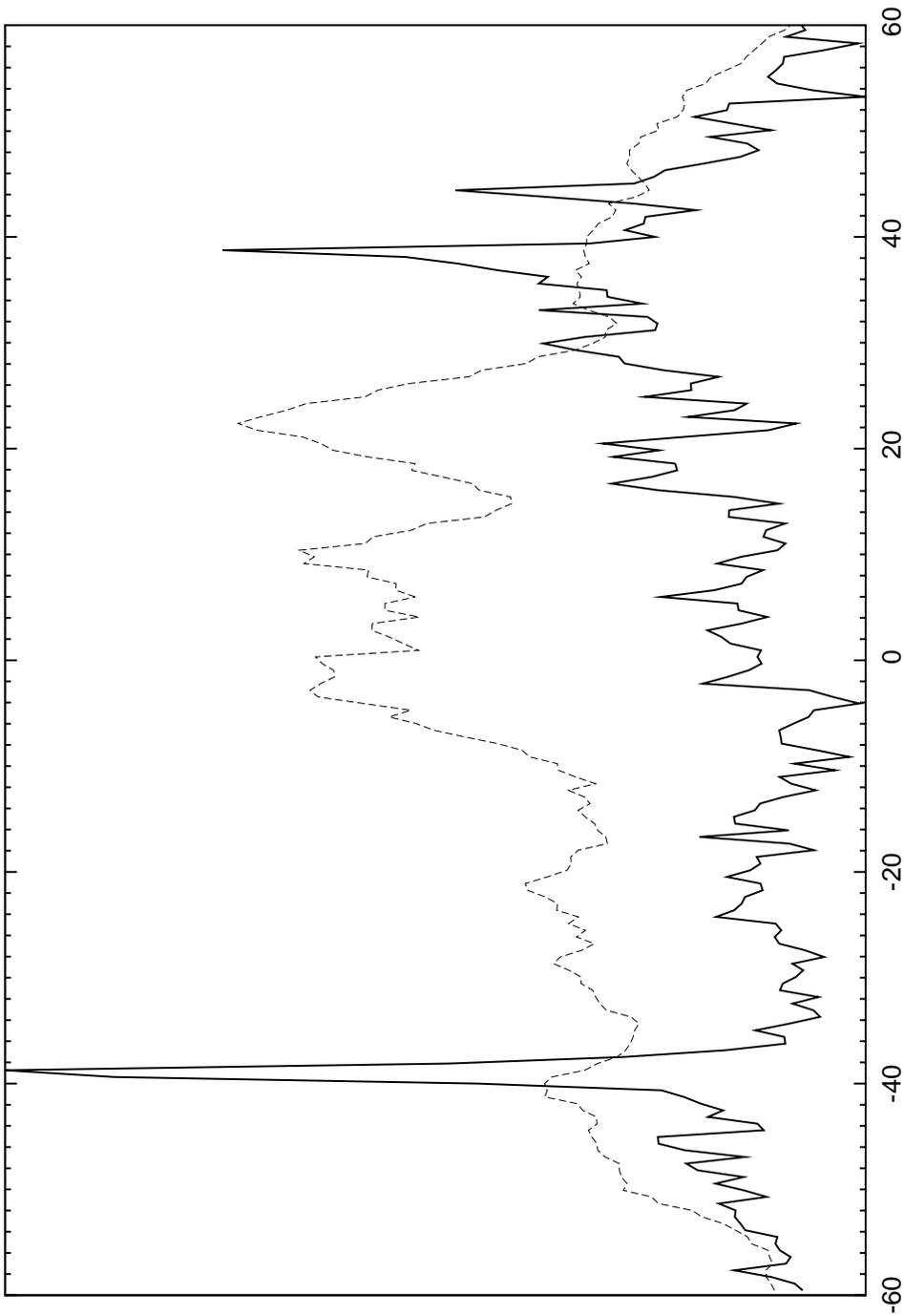}
\caption{Radial profile of \snrb.  The solid line represents \ha\ and the dashed line represents 0.35--0.85 keV smoothed X-rays.  X-axis units are parsecs. The origin corresponds to  05 06 01.87, -65 41 29.43.  The northeastern spur is visible near -45. }\label{rad0506}
\end{figure}

\snrb\ (\object{DEM L 72}, \object[\[FHW95\] B0505-6548]{FHW95 B0505-6548}) shows a clear shell structure in the optical data with \sii/\ha\ between 0.5 and 0.7. It is isolated from any other diffuse optical or X-ray emission.  It is roughly circular along the north edge, but flattened along the eastern side.  The southwestern edge shows a bright flat region with complex filamentary structure.  There appears to be a gap in the shell along the southern edge but closer investigation shows faint optical filaments extending southwestward and reconnecting with the flat region mentioned above.  Complex filamentary structure is visible inside the shell.  We also see a filament extending northeast from the remnant.  The fainter filaments are most clearly visible in \oiii\ images.  The optical shell and X-ray emission cover $7.3'$ or 110~pc along a northwest-southeast line, and $6.7'$ or 100~pc along a northeast-southwest line.  A northeastern spur extends a projected distance of $1.2'$ or 18~pc east of the main shell.  For the purposes of our calculations, we assign a definite size to the remnant by fitting an ellipse with radii of $2\farcm 7$ and $4\farcm 0$, or 40~pc and 60 pc, to the shell.  This ellipse encompasses the main body of the shell, along with the northeastern spur and the faint filament bounding the southern edge.  The shell is filled with diffuse soft X-ray emission but it is not as evenly distributed as in \snra .  Several local maxima are observed.  The X-ray emission seems to divide itself into three regions, one in the northwest corner, a second within the southern bulge, and a third along the eastern side, fully contained by the northeastern spur.  The emission from this last region seems to be generally fainter than that from the other two regions.  See Figures \ref{fig0506} and \ref{rad0506}.

We detected emission across the face of \snrb\ in the echelle spectra, although the pattern is not smooth.  Rather it shows several areas of convergence toward or deviation from systemic velocity.  The systemic component is distinguished from the telluric OH line.  Maximum inferred shell velocities are $\pm$ 90 \kms.

In the XMM data, we defined a circular source region covering the SNR and extracted the remainder of the FOV as the background.  After light curve filtering we were left with 9 ks of integration time.  We found a best fit temperature for the local hot bubble as $kT=0.09$~keV and for the Galactic halo as $kT=0.2$~keV, consistent with the results from the \snra\ background.  The best fit column density is 3.2$\times10^{21}$~cm$^{-2}$.  Using these background parameters, we fit our source region with an additional thermal plasma and found a gas temperature of $kT=0.17$~keV.

\subsection{\snrc}

\begin{figure}
\plotone{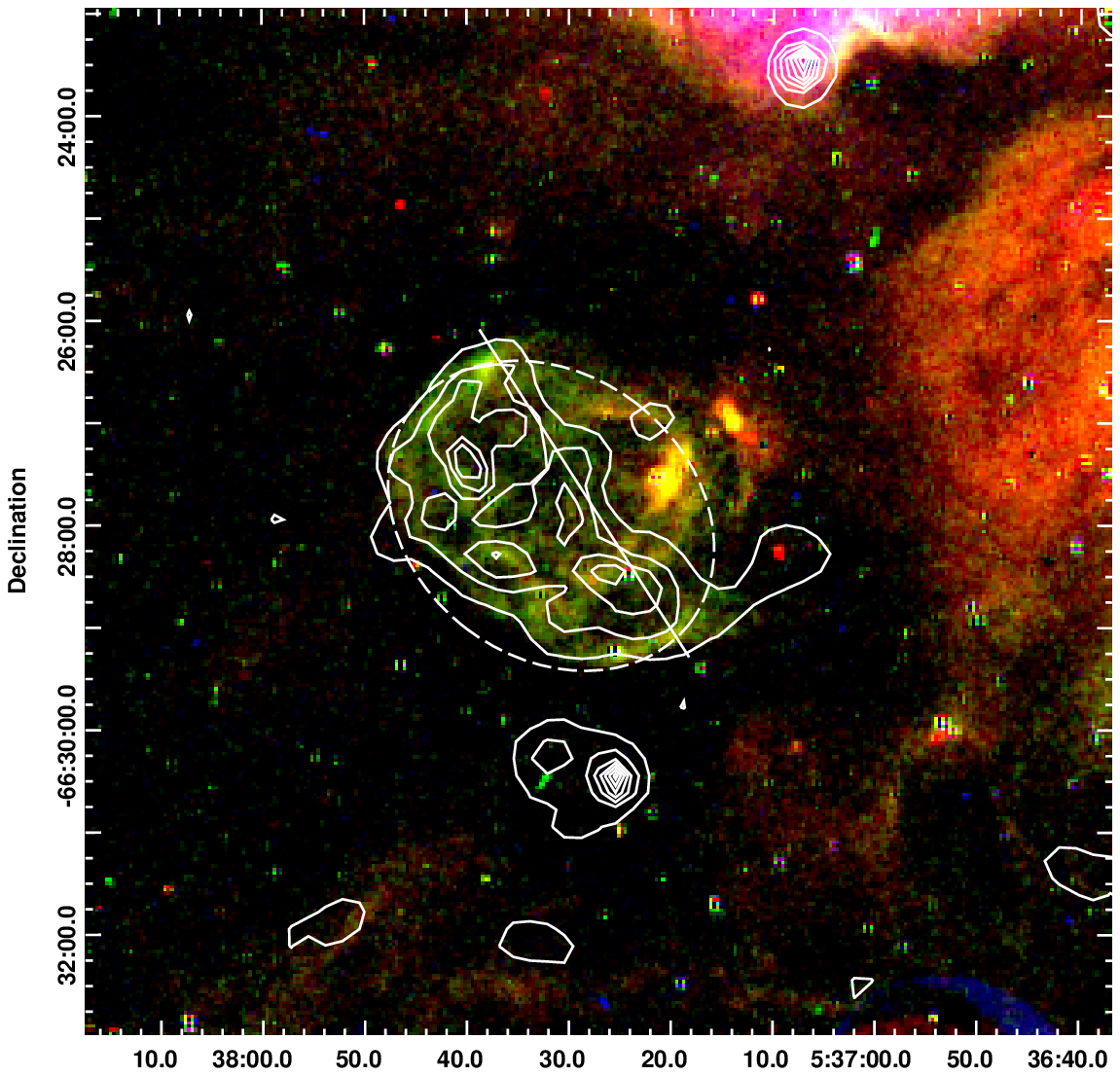}
\caption{Three-color image of \snrc\ with X-ray contours overlaid. Red is \ha, green is \sii, and blue is \oiii.  Contours are at levels of 2 counts per second per square degree in the energy band 0.35--0.85 keV.  The dashed ellipse corresponds to the SNR extent considered in the calculations of the X-ray and optical shell properties.  The line corresponds to the radial slice shown in Figure \ref{rad0537}.}\label{fig0537}
\end{figure}

\begin{figure}
\epsscale{1.5}
\plotone{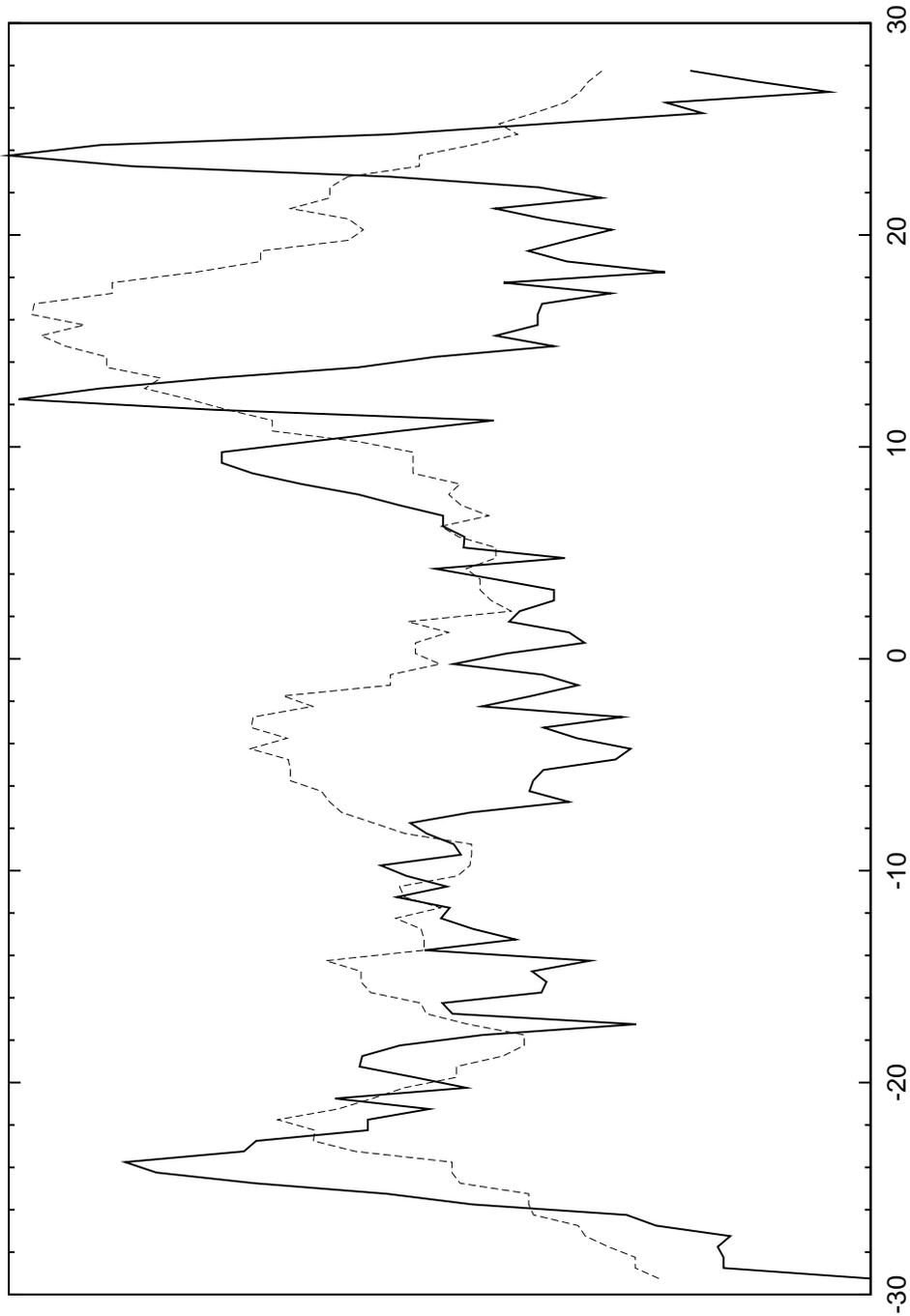}
\caption{Radial profile of \snrc.  The solid line represents \ha\ and the dashed line represents 0.35--0.85 keV smoothed X-rays.  X-axis units are parsecs. The origin corresponds to  05 37 28.49, -66 27 41.47.  The slice was chosen to lie along the western extent of the outer shell. }\label{rad0537}
\end{figure}

\snrc\ (\object{DEM L 256}) also lies projected near the \hii\ regions \object{DEM  L 251}, \object[DEM  L 253]{253}, and \object[DEM  L 264]{264}.  Optical filaments take the form of two nested shells.  The inner shell is complete and has a radius of $0.8'$ or 11.6~pc.  The outer shell covers roughly $180^\circ$ on the southeast half of the remnant with a radius of $1.6'$ or 23~pc  along its western extent and $1.3'$ or $20$~pc in the perpendicular direction.  Two clumps of \ha\ emission appear on the western side of the remnant.  One clump lies along the inner shell, and the second is roughly co-circular with the outer shell.  Our \sii/\ha\ ratio maps reveal that these clumps have ratios of $<0.4$, similar to the nearby \hii\ regions and in contrast to the values $>0.7$ found in the shell filaments.  Thus it seems that these clumps are not actually associated with the SNR.  The X-ray emission is confined to the eastern half of the remnant corresponding to the extent of the outer \ha\ shell and appears to be contained primarily between the two filaments.  See Figures \ref{fig0537} and \ref{rad0537}.

This SNR shows clear blue- and redshifted expansion patterns across the face, although the redshifted emission is considerably stronger than the blueshift counterpart.  A wide band of \ha\ emission is also visible from the surrounding \hii\ region.  The expansion pattern has two distinct sections, with distinct central velocities but equal expansion velocities of $\pm$ 55 \kms.  The break between the two patterns corresponds to the point where the echelle slit crosses the inner optical shell.  Strong \ha\ emission is seen converging to the systemic velocity at the location of the outer half-shell.  A corresponding convergence on the other end of the slit is much fainter, but this end of the slit does not intersect any part of the outer half-shell.

There is diffuse soft X-ray emission around the edge of the XMM field of view in the direction of \object{DEM  L 256}.  In this exposure, 20 ks of integration time were useable.  We extracted a background annulus around the SNR that is small enough to avoid the emission associated with the \hii\ regions.  Here we found best-fit temperatures of $kT=0.17$~keV for the local hot bubble and $kT=0.22$~keV for the Galactic halo, again consistent with the above findings.  Column density is fit as 5$\times10^{21}$~cm$^{-2}$.  Fitting the center of the annulus with this background plus a source plasma gave a gas temperature of $kT=0.22$~keV.

\section{Discussion}

The \sii/\ha\ ratio technique has been used before to discover SNRs which were too faint to be detected in previous surveys, for example, \citet{rosa04} and \citet{stupar}.  This technique when applied to the detailed and deep MCELS images turned up yet more SNR candidates which had been missed in previous optical studies\footnote{See for example Williams et al. 2010 (in preparation) and references therein,  online at http://hoth.ccssc.org/mcsnr/ }.  Using the thermal source models fitted to the X-ray spectra, we can derive a flux which can be converted to an estimated \rosat\ High Resolution Imager count rate using the PIMMS tool from HEASARC.  We found that \rosat\ would have collected roughly $1.3\times10^{-4}$ counts per second per square arcminute from \snrb\ and $3\times10^{-4}$ from \snrc\ over its primary energy band of 0.2--2.0 keV.  We were not able to calculate a flux for \snra\ since its spectrum was of too poor quality to achieve a satisfactory fit.  However we can set an upper limit on its flux by assuming that all emission from the source region is due to a single absorbed thermal plasma.  In this way we found a flux of $5.0\times10^{-14}$~erg s$^{-1}$ cm$^{-2}$ corresponding to $2.1\times10^{-3}$ counts per second per square arcminute for \rosat.  We found that the two remnants presented here for which we have good flux measurements are indeed fainter than all \rosat\ detected LMC remnants from the catalog of \citet{rosa99}.  Our upper limit for \snra\ is higher than three of the \rosat\ SNRs, but among those, one benefitted from a lengthy 109 ks exposure, and the other two detections are of too poor quality to allow a discussion in that reference.  We also note that \snrc\ coincides with \object[\[HP99\] 344]{HP 344}, although that object was not identified as a SNR candidate \citep{hp99}.  Again using PIMMS to convert our measured flux to a \rosat\ PSPC count rate over the range 0.1--2.4 keV we found an expected rate of $1.6\times10^{-2}$ counts per second, which compares favorably with the rate of $(1.4\pm0.2)\times10^{-2}$ counts per second from \citet{hp99}.

Our technique can not be expected to identify the complete population of faint SNRs, however.  As discussed in \citet{chu97} and \citet{cm90}, SNRs in superbubbles cleared out by OB associations lack the optical shell signature.  Several SNRs have been discovered within LMC superbubbles (SBs) solely by their X-ray emission in \rosat\ mosaics.  Based on the success of the present project in identifying previously unknown faint remnants, it seems likely that more such remnants exist, but may be lacking in optical signatures.  A deep X-ray survey of SBs with modern instruments would be necessary to fill out the population of LMC remnants.  Since the spectra of SNRs soften as they age, instrumentation with enhanced sensitivity to low energy photons would be especially beneficial.

Superbubbles in the LMC are often in excess of 100 pc, Êso the comparatively small sizes of these objects, as well as the lack of interior OB associations, argues for their formations as single SNRs, rather than as SBs. ÊLikewise, SBs tend to have fairly slow expansion velocities, close to those of the surrounding ISM. ÊThe derived expansion velocities are on the low side for SNRs, but are higher than one would expect for a SB \citep{SB}. ÊOnly \snrb\ is of comparable size to these SBs but has a higher expansion velocity, which argues for an SNR origin. ÊÊ\snra\ and \snrc\ Êare on the high side of velocity for SBs, are notably smaller than typical SBs, and do not have any known OB associations, which suggests that they also are likely SNRs.

We can estimate the density of the ambient ISM by summing the masses of the warm shell and hot gas and dividing by the volume of the remnant.  Technically one should also subtract the mass of the progenitor star, but in all three cases here the SNR masses are now much greater than a typical progenitor mass of $\sim$10M$_\odot$.  Moreover, this ISM density estimate will be strongly affected by the same approximations mentioned in \S2 with respect to calculating the shell mass, so these are at best a rough estimate.  For \snra, we were unable to fit a model to the source plasma and so we do not have information on the density of the hot gas.  In this case, we included only the mass of the shell and calculated a lower limit of $\sim4$ cm$^{-3}$ on the ambient density.  Even this lower limit is significantly higher than the densities of 0.6 and 1.4 cm$^{-3}$ which we estimated for \snrb\ and \snrc\ respectively.  This is consistent with the observation that \snra\ seems to be embedded in an \hii\ region.


An additional motivation for studying \snra\ was its close projection to the ring
nebula around the Wolf-Rayet star \object[\[M2002\] LMC 5735]{Br2}.  This ring nebula is unusual in its
bright nebular \heii$\lambda$4686 line emission, indicating very high
excitation \citep{garnett}.  Two other examples of \heii-emitting
nebulae are known in the LMC: the \heii\ nebula around the X-ray binary \object{LMC X-1}
\citep{pange} and \object[LHA 120-N 44C]{N44C} which has been suggested to be
ionized by the transient \object{LMC X-5} \citep{pmotch}.  Later studies demonstrated that the \heii\ and \ha\ line widths in the ring nebula are too narrow to be caused by dynamical interaction with the expanding SNR shell \citep{chu99}.  However, the possibility of excitation by an X-ray source within the ring \citep[e.g.,][]{chu00} or within the SNR had not been ruled out.  Our observations found no significant compact X-ray emission at the location or \object[\[M2002\] LMC 5735]{Br2}.  Furthermore, we see a smooth distribution of soft emission within the SNR shell with no indications of any compact or hard source.


The very irregular shell morphology of \snrb\ makes it difficult to assign it an exact size.  Measured at the extreme extents of the optical shell, it has radii of 50 $\times$ 55 pc which would make \snrb\ the largest known SNR in the LMC, larger even than \object{SNR 0450-70.9} \citep{rosa04}.  However, for the purposes of our quantitative estimates, we take a shell size of 40 $\times$ 60 pc, as mentioned above.  The brightest X-ray emission is seen in the western half of the SNR.  Here the optical shell flattens and becomes thicker and more filamentary.  Many optical clumps are also seen in this half of the shell.  This is consistent with the SNR encountering a region of higher ISM density which is being shocked by the blast wave.  As this material passes through the shock, it evaporates producing diffuse X-ray emission.  The faint optical filament along the extreme eastern edge of the remnant coincides with a line of X-ray emission.  This shock front may still be capable of heating the gas it encounters to X-ray emitting temperatures.  Elsewhere along the northern rim we see less emission along the shell and, in some areas, a gap between the optical shell and interior X-ray emission, suggesting that the shock front here is no longer powerful enough to fully ionize new gas and that previously heated gas has now cooled to below detectable levels.  

The expansion velocity of 90 \kms\ implies an age of $\sim$150 kyr.  Solving for $E/n$ as described in \S2 gives $E/n=2.6\times10^{51}$~erg~cm$^3$, in excellent agreement with the values calculated from the H$\alpha$ surface brightness and X-ray spectrum as listed in Table \ref{snrtable}.  However, such slow-moving shocks would not be capable of heating new gas to X-ray emitting temperatures.  Post-shock velocities of $\sim$140 \kms\ are needed to heat gas to a sufficient temperature to radiate strongly at 0.2 keV.  The evidence for newly shock-heated gas seen in the X-ray images may indicate that for at least some portions of the SNR front, the velocity is well in excess of 90 \kms.  If we instead consider the expansion velocity to be 140 \kms, we find an age of 95 kyr.  This is an issue seen in other remnants, e.g. \object[LHA 120-N 86]{N86} and \object[LHA 120-N 11L]{N11L}.  It may be that the optical material  with the greatest Doppler shifts has too low surface brightness for detection in echelle spectra.   In any case, the observed gas velocities are well below those expected at the end of the Sedov-Taylor phase, and thus the age of 150 kyr is likely an overestimate.  For comparison, an SNR of this radius but with gas velocity corresponding to the end of the Sedov-Taylor phase would give an age of $\sim$80 kyr.

When applying the simulated spectrum, we found a best fit at an age of 80 kyr.  However, the simulations seem to underpredict the prominent emission lines around 0.6 keV.   This could indicate that we have underestimated the abundances throughout our fitting process.  Returning to the background model and allowing the local hot bubble abundance to float, we found a best fit when $Z_\mathrm{LHB}=1.9$.  Applying the background model to the source spectrum we again found a best fit to the SNR emission when the source plasma abundances are set to 0.5.  Plugging in the simulated spectrum we found that an age of 100 kyr is now preferred, in better agreement with the analytically derived values.  However, one might expect that such an old remnant would be heavily dominated by the swept-up medium, and therefore show abundances close to those of the LMC.  Furthermore, such an old remnant will have evolved past the Sedov-Taylor phase which ends at velocities

It is worth pointing out that the physical parameters which we derived for this SNR are broadly similar to those found by \citet{rosa04} for \object{SNR 0450-70.9}.  Prior to that study, no interior X-ray emission had ever been detected from a remnant of this size.  Moreover, no remnant of this size is known within our galaxy.  Thus, a detailed study of two such advanced SNRs represents a significant step forward in our understanding of the long-term fate of these objects.  We also note that $P_\mathrm{hot}$ is significantly greater than $P_\mathrm{shell}$ for this remnant, suggesting some pressure-driven expansion even at these late ages.


X-ray emission from \snrc\ appears to trace the \ha\ filaments, again suggesting that this gas is recently shock heated.  The appearance of two distinct shells along with two expansion patterns in the echelle spectrum which correlate with the optical shells suggests a bilobed structure for the SNR.  The near edge of the SNR seems to be encountering denser material as evidenced by the X-ray emission corresponding to that shock wave and the stronger \ha\ emission converging to systemic velocity on that edge.  

The observed expansion velocity of 55 \kms\ is quite low and gives an age of 125 kyr.  The corresponding ratio $E/n=6\times10^{49}$~erg~cm$^3$ does not agree as well with the observed value of $1.6\times10^{50}$~erg~cm$^3$.  However, the thermal energy of the hot gas was estimated assuming filling of an entire ellipsoidal shell, whereas we see X-ray emission from only the eastern half.  Furthermore, if this is a bilobed structure which is not aligned with the line of sight, we are only measuring a projected component of the true expansion velocity.  Both of these considerations would bring the two numbers into closer agreement.  We calculate that the viewing angle of the bilobed expansion would need to be $\sim40^\circ$ in order for the two numbers to agree.  We also note that if the observed elliptical region of X-ray emission were actually a circular disc, a viewing angle of $\sim55^\circ$ would cause it to appear as it does.  This viewing angle would imply a true expansion velocity of 97 \kms\ which corresponds to an age of 70 kyr.
However, if the remnant is indeed encountering denser material, it may have undergone a more abrupt (non-adiabatic) deceleration, and again, this expansion velocity is outside the expected range of the Sedov-Taylor phase, netting a younger age for a given observed radius and expansion velocity.

The uncertainty in the properties of this SNR makes it difficult to choose accurate parameters for the simulations.  Moreover, our simulations are spherically symmetric, which is clearly not the case for this remnant.  Simulations run over a range of reasonable parameters all result in fits with ages $< 50$~kyr, lending support to the above consideration.

\section{Conclusions}

We have presented a multi-wavelength study of three recently discovered SNRs in the LMC, featuring new \xmm\ observations.  We fit background spectra from each observation with a detailed model including soft proton contamination, instrumental lines, the local hot bubble, and extragalactic components.  These fits gave consistent values for the background parameters across the three observations.  We then fit source spectra with  thermal plasma models, and compare the fits to simulated spectra.  We analytically estimated the physical properties of the SNRs.

\snra\ is confirmed as an SNR due to its [SII]/Ha ratio, velocity expansion pattern, and soft X-ray emission, although we cannot fully characterise that emission.  \snrb\ appears from our findings to be a large, relatively old SNR,  similar to SNR 0450$-$709.   It has fairly disorganized expansion and X-ray emission that appears to be due to a combination of recently shocked material and older ``fossil'' radiation.  \snrc\ appears to be encountering higher density material to one side and shows signs of a bipolar expansion geometry.  Although we were able to obtain good fits to its spectrum, the age is difficult to determine precisely due to uncertainties in its geometry and environment.



\acknowledgments

The authors would like to acknowledge David Henley for helpful comments and assistance in the conversion of the simulated spectra to Xspec models.

Cerro Tololo Inter-American Observatory (CTIO) is operated by the Association
of Universities for Research in Astronomy Inc. (AURA), under a cooperative agreement with the National
Science Foundation (NSF) as part of the National Optical Astronomy Observatories (NOAO). We gratefully acknowledge the support of CTIO and all the assistance which has been provided in upgrading the Curtis Schmidt telescope. The MCELS is funded through the support of the Dean B. McLaughlin fund at the University of Michigan and through NSF grant 9540747.




\clearpage



\clearpage

\clearpage

\clearpage

\clearpage

\clearpage

\clearpage

\clearpage

\clearpage

%

\begin{deluxetable}{lcccc}
\tablecaption{\xmm\ Observation Log \label{obslog}}
\tablewidth{0pt}
\tablehead{
\colhead{Obs ID}                 & 
\colhead{RA}               &
\colhead{Dec.}                & 
\colhead{Start Time}		&
\colhead{Exp. Time (ks)}                
}
\startdata
205260101	& 04h49m40.00s	& -69$^\circ$21'49.0''	& 2004-02-24 16:19:07	&18.7	\\
205260201	& 05h06m05.00s	& -65$^\circ$41'30.0''	& 2004-01-16 18:06:28	&16.9	\\
301410601	& 05h37m32.00s	& -66$^\circ$28'26.6''	& 2005-10-02 09:33:11	& 21.8	\\

\enddata
\end{deluxetable}

\clearpage

\begin{deluxetable}{lccc}
\tablecaption{Physical Properties of SNRs.  The X-ray fit parameters are from an APEC model. \label{snrtable}}
\tablewidth{0pt}
\tablehead{
\colhead{}                 & 
\colhead{\snra}               &
\colhead{\snrb}                & 
\colhead{\snrc}                
}
\startdata

$R$~(pc)	& 19$\times$11	& 60$\times$40	& 24$\times$20	\\

SB~($10^{-15}$)	& 0.9		& 0.5		& 0.4		\\

$v_{\mathrm{exp}}$~(km~s$^{-1}$)	& 70	& 90	& 55		\\

$n_{\mathrm{e,shell}}$~(cm$^{-3}$)	& 5.5(2)		& 2.99(8)	& 3.30(4)		\\

$V_{\mathrm{shell}}$~(cm$^{3}$)	& $1.4\times 10^{59}$	& $1.7\times 10^{60}$	& $3.2\times 10^{59}$	\\

$M_{\mathrm{shell}}$~(\Msun)	& 820(30)	& 5.6(1)$\times10^3$	& 1.14(1)$\times10^3$	\\

$K$~(erg)	& $4(1)\times 10^{49}$	& $4(1)\times 10^{50}$	& $3(1)\times 10^{49}$	\\

$P_{\mathrm{shell}}$~(dyne~cm$^{-2}$)	& $1.51(5)\times 10^{-11}$	& $8.2(2)\times 10^{-12}$	& $9.1(1)\times 10^{-12}$	\\

$kT$~(keV)	& 	& 0.17(1)	& 0.22(5)	\\

Normalization ($10^{-14}$ cm$^{-5}$)	&	& 0.004(2)	& 0.0009(9) \\

$V$ (cm$^3$)	&	& $1.3(3)\times10^{61}$	& $1.0(1)\times10^{60}$ \\

$n_{\mathrm{e,hot}}$~(cm$^{-3}$)	& 	& 0.11(4)	& 0.2(2)	\\

$M_{\mathrm{hot}}$~(\Msun)	& 	& $1.3(9)\times10^3$	& 2(2)$\times10^2$	\\

$E_{\mathrm{th}}$~(erg)	& 	& $1.1(8)\times 10^{51}$	& $2(2)\times 10^{50}$	\\

$P_{\mathrm{hot}}$~(dyne~cm$^{-2}$)	& 	& $5(2)\times 10^{-11}$	& $1(1)\times 10^{-10}$	\\

$n_\mathrm{ISM}$~(cm$^{-3})$	& $>4.0$	& 0.5	1(1)& 1.40(2)\\

\enddata
\end{deluxetable}

\clearpage

\begin{deluxetable}{lcccc}
\tablecaption{Background Spectral Parameters \label{bgtable}}
\tablewidth{0pt}
\tablehead{
	\colhead{}                 & 
	\colhead{\snra\ }               &
	\colhead{\snra\ }		&
	\colhead{\snrb}                & 
	\colhead{\snrc}      \\
	\colhead{}				&
	\colhead{\hii\ region}          &
	\colhead{clear region}	
}
\startdata
Soft Proton Index										& 1.9(6)					& 1.9(6)			& 1.0(1)	& 1.39(7) \\ \\

Soft Proton Norm. 			& 0.02(2)					& 0.003(3)			& 0.04(1)	& 0.048(5) \\

(cts keV$^{-1}$ cm$^{-2}$ s$^{-1}$) \\ \\ 

Local Bubble Temp. (keV)								& 0.11(1)					& 0.11(1)			& 0.09(1)	& 0.17(4) \\ \\

Local Bubble Norm. ($10^{-14}$ cm$^{-5}$)					& 4(4)$\times10^{-4}$			& 7(7)$\times10^{-4}$	& 0.002(2)	& 3(1)$\times10^{-4}$ \\ \\
$\NH$~($10^{22}$~cm$^{-2}$)	& 0.37(7)	& 0.37(7)	 & 0.32(7)	& 0.5(1)	\\ \\

Extragal. Thermal Temp. (keV)								& 0.27(4)			& 0.27(4)			& 0.20(2)	& 0.22(2) \\ 
	& 0.19(4) \\

Extragal. Thermal Norm. 				& 7(7)$\times10^{-4}$ 	& 0.001(1)		& 0.003(2)	& 0.005(3) \\
($10^{-14}$ cm$^{-5}$) & 0.02(1) \\ \\

Extragal. Norm.  	& 2(2)					& 2.1(7)			& 3(2)	& 0.9(9) \\
(10$^{-4}$ cts keV$^{-1}$ cm$^{-2}$ s$^{-1}$) \\

\enddata
\end{deluxetable}




\end{document}